\documentclass[a4paper]{spie}

\usepackage{amssymb}
\usepackage{amsmath}
\usepackage[]{graphicx}
\newcommand{\benumerate}{\begin{enumerate}}
\newcommand{\eenumerate}{\end{enumerate}}

\newcommand{\bitemize}{\begin{itemize}}

\newcommand{\eitemize}{\end{itemize}}
\newcommand{\der}[2]{\frac{\partial #1}{\partial #2}}

\newcommand{\dermixd}[3]{\frac{\partial^{2} #1}{\partial #2 ~\partial #3}}

\newcommand{\simbon}[2]{\mathop{\rm #1}\limits_{#2}}

\newcommand{\paperspie}[7]{#1, ``#2'', {\it #3}, {\bf #4}, pp. #5-#6, #7}
\newcommand{\bookspiech}[6]{#1, {\it #2}, #3, #4, #5, #6}
\newcommand{\bookspie}[5]{#1, {\it #2}, #3, #4, #5}

\title{Paraxial light in a Cole-Cole nonlocal medium: integrable regimes
 and singularities.}

\author{Boris Konopelchenko and Antonio Moro\supit{$\ast$}
\skiplinehalf Dipartimento di Fisica dell'Universit\`{a}
di Lecce \\
and INFN, Sezione di Lecce, via Arnesano I-73100 Lecce, Italy}

\authorinfo{\supit{$\ast$}For correspondence contact E-mail address:
 antonio.moro@le.infn.it}

\begin{document}
\maketitle

\begin{abstract}
Nonlocal nonlinear Schr\"odinger-type equation is derived as a
model to describe paraxial light propagation in nonlinear media
with different `degrees' of nonlocality. High frequency limit of
this equation is studied under specific assumptions of Cole-Cole
dispersion law and a slow dependence along propagating direction.
Phase equations are integrable and they correspond to
dispersionless limit of Veselov-Novikov hierarchy. Analysis of
compatibility among intensity law (dependence of intensity on the
refractive index) and high frequency limit of Poynting vector
conservation law reveals the existence of singular wavefronts. It
is shown that beams features depend critically on the orientation
properties of quasiconformal mappings of the plane. Another class
of wavefronts, whatever is intensity law, is provided by harmonic
minimal surfaces. Illustrative example is given by helicoid
surface. Compatibility with first and third degree nonlocal
perturbations and explicit solutions are also discussed.

\end{abstract}


\section{Introduction.}
\label{sec_intro}

Propagation of the light through spatially nonlocal media is one
of main subjects in modern nonlinear
optics\cite{Segev1,Segev2,Snyder,Assanto1,Assanto2,SegevRev}.
Recently, it has been shown, both theoretically\cite{Snyder} and
experimentally\cite{Assanto1,Assanto2}, that highly nonlocal
Kerr-type media supports propagation of stable laser beams,
usually referred to as ``solitons''. Laser beams propagating in
Kerr-type media are modelled by the celebrated ($2+1$)D nonlinear
Schr\"odinger (NLS) equation. Unlike ($1+1$)D NLS equation, it is
not integrable by inverse scattering method\cite{Ablowitz}. In the
case of high nonlocal media, (2+1)D NLS equation can be
approximated by quantum harmonic oscillator (QHO) equation. Using
results concerning QHO in the optics context, it has been possible
to show the existence of solitons\cite{Snyder}. Nevertheless, a
complete understanding of the r\^ole played by spatial nonlocal
effects in the propagation of radiation in nonlinear media is
still an open problem. Hence, it is of interest to investigate
possible integrable regimes for three-dimensional nonlinear
models.

In the present paper, we propose the nonlocal nonlinear
Schr\"odinger-type (NNLS) equation as the model to describe a beam
propagating in a medium with a certain spatial nonlocality degree
and obeying some intensity law. By ``intensity law'' we mean
certain functional dependence among dielectric function and
intensity of the beam. We study the high frequency limit under
specific assumptions, namely, the Cole-Cole dispersion law for the
medium and suitably slow variation of the fields along propagating
direction. Separating behaviors on the transverse and propagating
directions, we derive two couples of equations for the phase and
the intensity. Due to the intensity law, phase and intensity
equations are strongly coupled and their compatibility appears to
be a highly nontrivial problem.

As far as phase equations concerned, the compatibility has been
studied in the papers~[\citenum{Moro1,Moro2}]. It has been shown
that compatibility condition corresponds to the request that
refractive index satisfies an infinite set of integrable equations
referred to as dispersionless Veselov-Novikov (dVN) hierarchy. In
particular, degrees of spatial nonlocality are in one-to-one
correspondence with the orders of equations of dVN hierarchy.

So, we study compatibility among phase and intensity equations for
certain assigned intensity law. We focus first on transverse
equations. Their compatibility leads us to the study of a second
order nonlinear partial differential equation. Since for most of
physical models it satisfies the ellipticity condition we restrict
ourselves to the class of so-called {\em elliptic intensity laws}.

The use of known results in the theory of elliptic equations and
the special properties of Beltrami equation allows us to conclude
that, for a generic elliptic intensity law, there exists singular
wavefronts which can be used to describe self-guided beams.
Moreover, we demonstrate that harmonic minimal surfaces provide us
with a class of wavefronts for any elliptic intensity law. As
explicit examples, helicoidal wavefronts are discussed. We stress
that the helicoid is a well known example of singular wavefront
which possesses several interesting
properties\cite{Bryngdahl,Berry,Vaughan,Baranova,Coullet,Soskin1,Soskin2}.
We show that helicoidal wavefronts are preserved for a specific
class of first degree nonlocal perturbations up to next-to-leading
order contributions. We present a simple example where the first
degree nonlocality is responsible of a slight stretching or
compression of the helicoid's pitch.

General problem of compatibility among equations for transverse
and propagating direction is difficult as much as crucial. In our
model the consideration of propagating direction contributions is
equivalent to inclusion of nonlocal terms. Besides, in the local
case, equations for propagating direction are automatically
compatible for any intensity law and it is sufficient to consider
the transverse equations. We demonstrate that there exists a class
of solutions for the third degree nonlocality which are in
agreement with a suitable intensity law. Exploiting
hydrodynamic-type reductions, found in the
paper~[\citenum{Bogdanov}], we present some explicit formulae.

The paper is organized as follows: in section~\ref{sec_NNLS}
derivation of nonlocal nonlinear Schr\"odinger equation is
presented. In the section~\ref{sec_GO} phase equations and high
frequency limit of Poynting vector conservation law in a Cole-Cole
medium are displayed. A study of compatibility among intensity law
and transverse equations and properties of solutions is presented
in the section~\ref{sec_transverse}. Compatibility of intensity
law and nonlocal perturbations is illustrated in
section~\ref{sec_symmmetry} by an explicit example. Some
concluding remarks close the paper.

\section{Nonlocal NLS-type equation.}
\label{sec_NNLS}
Let us consider the nonlinear Maxwell equations
\begin{gather}
\label{Maxwell_nonlinear}
\begin{aligned}
\nabla \wedge {\bf E} + \der{{\bf B}}{t} &= 0 \quad{}\quad{}\nabla
\cdot
{\bf D} = 0  \\
\nabla \wedge {\bf B} - \der{{\bf D}}{t} &= 0 \quad{}\quad{}
\nabla \cdot {\bf B} = 0
\end{aligned}
\end{gather}
where the displacement vector ${\bf D} = {\bf D}({\bf E})$ depends
nonlinearly on the electric field. We look for time-oscillating
solutions such that
\begin{equation*}
{\bf E}\left ({\bf x}, t \right) = {\bf E}\left({\bf x} \right)
e^{- i \omega t}, \quad{} {\bf D}\left ({\bf x}, t \right) = {\bf
D}\left({\bf x} \right) e^{- i \omega t}.
\end{equation*}
Under these assumptions one gets the following second order
equation\cite{Born}
\begin{equation}
\label{Maxwell_second} \nabla^{2} {\bf E} + \omega^{2} {\bf D} -
\nabla \left(\nabla \cdot {\bf E} \right) = 0.
\end{equation}
Displacement vector ${\bf D}$ is chosen in a way to collect both
nonlinear and nonlocal effects, i.e.
\begin{equation}
\label{displacement} {\bf D} = \varepsilon {\bf E} + {\bf
F}_{NLoc}.
\end{equation}
First term of the r.h.s. contains the nonlinearity. Dielectric
function $\varepsilon$ is assumed to contain a weakly nonlinear
perturbation of the form
\begin{equation}
\varepsilon  = \varepsilon_{0} + \alpha^{2} \tilde{\varepsilon},
\quad{}\alpha << 1,
\end{equation}
 where $\tilde{\varepsilon}$ depends on the intensity of
 the radiation.

The second term in the r.h.s. of~(\ref{displacement})  provides us
with the spatial nonlocal response of the medium. We assume it to
be of the form
\begin{equation}
\label{nonlocal_term} {\bf F}_{NLoc} = \alpha \left(\sum_{n=1}^{3}
c_{n} \der{{\bf E}}{x_{n}}  + \sum_{l,m=1}^{3} c_{lm}
\dermixd{{\bf E}}{x_{l}}{x_{m}} + \dots \right).
\end{equation}
One gets the expression~(\ref{nonlocal_term}) for the nonlocal
term considering an integral relation among the displacement and
the electric field where the kernel is a sum of $\delta$-functions
and spatial derivatives. We define the degree of nonlocality as
the maximum derivative order which appears in
expression~(\ref{nonlocal_term}).

\noindent Paraxial approximation is performed according to the
rule (see e.g. Ref.~\citenum{Landau})
\begin{equation}
\label{rule_paraxial} {\bf E} \rightarrow \alpha {\bf E} e^{i k z}
\quad{} {\bf D} \rightarrow \alpha {\bf D} e^{i k z}
\end{equation}
(setting vacuum light speed $c = 1$, $k = \sqrt{\varepsilon_{0}}
\omega$), and the amplitudes are assumed to depend slowly on the
coordinates in such a way that the gradient acts as follows
\begin{equation}
\label{rule_gradient} \nabla = \left(\der{}{x},\der{}{y},\der{}{z}
\right) \rightarrow \left(\alpha \nabla_{\bot},\alpha^{2}
\der{}{z} \right).
\end{equation}
Using the expression~(\ref{displacement}) and the
rules~(\ref{rule_paraxial}) and~(\ref{rule_gradient}) in the
second order equation~(\ref{Maxwell_second}), one shows that the
first nontrivial contribution is of the order ${\cal
O}\left(\alpha^{3} \right)$ and it provides us with the nonlocal
nonlinear Schr\"odinger-type (NNLS) equation
\begin{equation}
\label{NNLS} 2 i \sqrt{\varepsilon_{0}} \omega \der{{\bf E}}{z} +
\nabla_{\bot}^{2}{\bf E} + \omega^{2} \tilde{\varepsilon} {\bf E}
+ {\cal F}_{NLoc} = 0
\end{equation}
where
\begin{equation}
F_{NLoc} = {\cal F}_{NLoc} + {\cal O}\left (\alpha^{4} \right).
\end{equation}
Note that in absence of nonlocal contribution ${\cal F}_{NLoc}$
and for a Kerr type nonlinear response of the medium,
equation~(\ref{NNLS}) becomes the standard nonlinear Schr\"odinger
equation in $2+1-$dimensions.

\section{Geometrical optics limit of NNLS equation in Cole-Cole media.}
\label{sec_GO} In this section we study the geometrical optics
limit of NNLS equation for a medium which satisfies so-called
Cole-Cole dispersion law\cite{ColeCole}
 \begin{equation}
 \label{ColeCole}
\tilde{\varepsilon} = \tilde{\varepsilon}_{0} +
\frac{\tilde{\varepsilon}_{1}}{1+ (i \tau_{0} \omega)^{2 \nu}},
\quad 0<\nu< \frac{1}{2},
 \end{equation}
where $\tau_{0}$ is the relaxation time and $\nu$ is a suitable
phenomenological exponent whose range of values is of crucial
importance in the following derivations. Relation~(\ref{ColeCole})
describes frequency dependence of dielectric function for several
polar media as well as liquid crystals and various biological
tissues (see e.g. Refs.~[\citenum{ColeCole,ColeRecent}]).

We study a particular high frequency regime of NNLS equation where
the phase of electric field varies slowly along $z$-direction in
such a way that
\begin{equation}
\label{slowder} z \rightarrow \omega^{2 \nu} z, \quad{} \der{}{z}
\rightarrow \omega^{- 2 \nu}\der{}{z}.
\end{equation}

We also assume nonlocal effects to be weak in high frequency limit
in such a way that they contribute at most with $\omega^{-2
\nu}-$order terms
\begin{equation}
\label{nonlocal_order} {\cal F}_{NLoc} = \omega^{-2 \nu}
\tilde{{\cal F}}_{NLoc} + {\cal O}\left (\omega^{-4 \nu} \right )
\end{equation}
In order to calculate geometrical optics limit we adopt, as usual,
the following representation of electric field
\begin{equation}
{\bf E} = {\bf f}(x,y, \omega^{2 \nu}) \ e^{i \omega
S(x,y,\omega^{2 \nu} z)}.
\end{equation}
Expansion of all functions with respect to small parameter
$\omega^{-2 \nu}$
\begin{gather}
\label{expansion}
\begin{aligned}
{\bf f}(x,y,\omega^{2 \nu} z) &= {\bf f}(x,y,z) + \omega^{- 2 \nu}
{\bf f}_{1} (x,y,z) + \dots \\
S(x,y,\omega^{2 \nu} z) &= S(x,y,z) + \omega^{- 2 \nu} S_{1}
(x,y,z) + \dots \\
\tilde{\varepsilon}(x,y,\omega^{2 \nu} z) &=
\tilde{\varepsilon}(x,y,z) + \omega^{- 2 \nu}
\tilde{\varepsilon}_{1} (x,y,z) + \dots
\end{aligned}
\end{gather}
together with dispersion law~(\ref{ColeCole}) and the
prescriptions~(\ref{slowder}) leads us to the high frequency limit
of NNLS equation at the orders $\omega^{2}$ and $\omega^{2-2
\nu}$:
\begin{subequations}
\label{phase}
\begin{align}
\label{eikonal}
\left( \nabla_{\bot} S \right)^{2} &= 4 u, \\
\label{z_dependence} S_{z} &= \varphi(S_{x},S_{y},x,y,z)
\end{align}
\end{subequations}
where $4 u = \tilde{\varepsilon}_{0}$ and $\varphi = \left (
\tilde{\varepsilon}_{1} + \tilde{{\cal F}}_{NLoc} \right) /\left(2
\sqrt{\varepsilon_{0}} \right)$. For convenience we denote the
partial derivative  of a function $f$ with respect to variable
$\eta$ as $f_{\eta}$. We stress that condition on the exponent $0
< \nu < 1/2$ is crucial to separate equation~(\ref{z_dependence})
from polarization contributions. Equation~(\ref{eikonal}) is the
standard eikonal equation in two-dimensions while the function
$\varphi$ in equation~(\ref{z_dependence}) is an $N$-degree
polynomial in $S_{x}$ and $S_{y}$ and it describes an $N$-degree
nonlocal response. Moreover, in agreement with inversion phase
symmetry ($S \to - S$) of the eikonal equation, polynomial degree
of function $\varphi$ must be odd. We note also that the system of
equations~(\ref{phase}) has been derived first directly from the
Maxwell equations\cite{Moro2}. In the present paper, we will
consider local response and first and third degree nonlocal
responses.

\noindent In local case we have
\begin{equation}
\label{local_hp} \varphi = \varphi (z), \quad{} u  = u (x,y),
\end{equation}
where $\varphi$ and $u$ are certain function of their arguments.
For a first degree nonlocality we have
\begin{equation}
\label{firstnonlocal_phi_hp} \varphi = \alpha_{1} S_{x} +
\alpha_{2} S_{y}, \quad{}
\end{equation}
and $u$ must satisfy the following linear equation
\begin{equation}
\label{firstnonlocal_u_hp} u_{z} = \left (\alpha_{1} u\right)_{x}
+ \left (\alpha_{2} u\right)_{y}
\end{equation}
where $\alpha_{1}$ and $\alpha_{2}$ are harmonic conjugate
functions (they satisfy Cauchy-Riemann conditions $\alpha_{1x} =
\alpha_{2y}$ and $\alpha_{1y} = - \alpha_{2x}$). Finally, for a
third degree nonlocal response we have
\begin{equation}
\label{thirdnonlocal_phi_hp} \varphi = \frac{1}{4} S_{x}^{3} -
\frac{3}{4} S_{x} S_{y}^{2} + V_{1} S_{x} + V_{2} S_{y}
\end{equation}
and $u$ satisfies the dispersionless Veselov-Novikov equation
\begin{gather}
\label{dVN}
\begin{aligned}
&u_{z} = \left (V_{1} u \right)_{x} + \left (V_{2} u \right), \\
&V_{1x} - V_{2y} = - 3 u_{x}, \\
&V_{1y} + V_{2x} = 3 u_{y}.
\end{aligned}
\end{gather}

In order to complete geometrical optics description we construct
high frequency limit of Poynting vector conservation law
\begin{equation}
\label{poynting} \nabla \cdot {\bf P} = 0.
\end{equation}
Poynting vector can be written down as follows
\begin{equation}
{\bf P} = I \frac{\nabla S^{\ast}}{\left | \nabla S^{\ast} \right
|}
\end{equation}
where
\begin{equation}
\label{total_phase} S^{\ast}  = \sqrt{\varepsilon}_{0} z +
S(x,y,\omega^{2 \nu}z)
\end{equation}
is the total phase of electric field and $I$ is the intensity. In
the following, for sake of simplicity, we set dielectric constant
$\varepsilon_{0} = 1$. Using the previous prescriptions to perform
high frequency limit in paraxial approximation and expanding
intensity
\begin{equation}
\label{intexpansion} I = I_{0} + \omega^{-2 \nu} I_{1},
\end{equation}
one concludes that equation~(\ref{poynting}) leads to the
following set of equations for $I_{0}$ and $I_{1}$
\begin{subequations}
\label{intensity}
\begin{align}
\label{I0}
\nabla_{\bot} I_{0} \cdot \nabla_{\bot} S + I_{0} \nabla_{\bot}^{2}S &= 0\\
\label{I1} \nabla_{\bot} I_{1} \cdot \nabla_{\bot} S + I_{1}
\nabla_{\bot}^{2} S +  \der{I_{0}}{z} &= 0.
\end{align}
\end{subequations}
In this regime, properties of the light are completely described
by the systems of equations~(\ref{phase}) and~(\ref{intensity}).
If no additional conditions are considered, once compatibility of
equations~(\ref{phase}) is guaranteed, linear
system~(\ref{intensity}) provides us with leading orders intensity
contributions $I_{0}$ and $I_{1}$.

Integrability of the phase equations~(\ref{phase}) has been
discussed in details in Refs.~[\citenum{Moro1,Moro2}]. In
particular, it has been shown that for a function $\varphi$ with a
polynomial dependence on $S_{x}$ and $S_{y}$, compatibility
condition of equations~(\ref{phase}) selects the class of
refractive indices which satisfy the dispersionless limit of the
Veselov-Novikov (dVN) hierarchy. In particular, dVN hierarchy
allows us to classify different degrees of nonlocal response by a
one-to-one correspondence among degrees of the polynomials
$\varphi$ and the equations of the hierarchy.

We mention that dVN hierarchy is amenable by the quasiclassical
$\bar{\partial}$-dressing method\cite{Moro2} and a reduction
method based on symmetry constraints appears to be an effective
method to construct exact solutions\cite{Bogdanov}.

\section{Phase and intensity transverse equations.}
\label{sec_transverse}
\subsection{Elliptic intensity law.}
Several models in nonlinear optics analyze a self-action of the
medium assuming that the influence of an external electromagnetic
field is described by a functional dependence among intensity and
refractive index $I = I(u)$. This is the form of intensity law in
the geometrical optics limit. Kerr type media, where $I \propto
u$, and saturable nonlinear media such that $u \propto \log (1 +
I/I_{t})$, where $I_{t}$ is the so-called threshold intensity, are
two famous examples. Of course, for self-action nonlinear media
models phase equations~(\ref{phase}) and intensity
equations~(\ref{intensity}) appear to be strongly coupled and the
study of their compatibility is a highly non-trivial problem. In
the present section, we analyze compatibility and properties of
equations~(\ref{eikonal}) and~(\ref{I0}), here referred to as
transverse equations. We note that this analysis is quite trivial
in the case of the media exhibiting local response. Indeed,
according to relations~(\ref{local_hp}) $u$ does not depend on the
$z$-variable. Then, equation~(\ref{I1}) coincides with
equation~(\ref{I0}) and they are automatically compatible
with~(\ref{z_dependence}).

Let us consider a self-action nonlinear medium obeying intensity
law $I_{0} = I_{0}(u)$. Observing that $\nabla_{\bot} I_{0} =
I_{0}' \nabla_{\bot} u$, where ``prime'' means the derivative with
respect its argument $u$,  and using equation~(\ref{eikonal})
in~(\ref{I0}), one gets the following second order partial
differential equation
\begin{equation}
\label{S_elliptic} A S_{xx} + B S_{yy} + 2 C S_{xy} = 0,
\end{equation}
where by definition ${\cal I}_{0} = \log I_{0}$ and
\begin{equation}
A = {\cal I}_{0}\;' S_{x}^{2} + 2, \quad{} B = {\cal I }_{0}\;'
S_{y}^{2} + 2, \quad{} C = {\cal I }_{0}\;' S_{x} S_{y}.
\end{equation}
Recall that properties of a second order partial differential
equation depend critically on the discriminant
\begin{equation}
\Delta = A B - C^{2}.
\end{equation}
If $A > 0$, equation is of the elliptic type when $\Delta > 0$ ,
parabolic when $\Delta = 0$ and hyperbolic when $\Delta < 0$. In
this paper we focus on the elliptic case motivated by the fact
that a wide class of such models for optical spatial solitons is
provided by a Kerr dependence or logarithmic saturable
nonlinearities. Indeed, one can simply verify that for a Kerr-type
intensity law of the form $I = u^{\beta}$ where $\beta > 0$ and a
logarithmic one, $u = \log (1 + I_{0}/I_{0,t})$, one has $\Delta =
4 (2 {\cal I}_{0}\;' u + 1)
> 0$ uniformly. Ellipticity of equation~(\ref{S_elliptic})
seems to be associated to important physical properties of a wide
class of nonlinear media. Moreover, elliptic equations such
as~(\ref{S_elliptic}) have been well studied in mathematics and
possess several remarkable analytical and geometrical
properties\cite{Bojarski,Iwaniec,Bers,Vekua,Ahlfors,Lehto}.

Now, we will study equation~(\ref{S_elliptic}). Introducing
complex variable $\lambda = x + i y $ and the complex gradient $w
= S_{x} - i S_{y}$, one can rewrite equation~(\ref{S_elliptic})
equivalently as follows
\begin{equation}
\label{S_elliptic_complex} a w_{\lambda} + b w_{\bar{\lambda}} +
\bar{a} \bar{w}_{\bar{\lambda}} + \bar{b} \bar{w}_{\lambda} = 0,
\end{equation}
where
\begin{align}
a = \frac{1}{2} \left(A - B + 2 i C \right), \quad{} b =
\frac{1}{2} \left (A + B - 2 i C \right).
\end{align}
A class of solutions of equation~(\ref{S_elliptic_complex}) is
provided by the solutions of, the so-called, nonlinear Beltrami
equation
\begin{equation}
\label{Beltrami_nlin} w_{\bar{\lambda}} = \mu \left (w, \bar{w}
\right) w_{\lambda}, \quad{} \mu = - \frac{a}{b}.
\end{equation}
It is straightforward to verify that ellipticity condition $\Delta
> 0$ implies $\left | \mu \right | < 1$. Beltrami equation describes so-called
quasiconformal mapping of the plane of complex dilatation
$\mu$\cite{Ahlfors,Lehto}. This fact highlights an intriguing
connection among geometrical optics and quasiconformal mapping
theory.

For instance, let us consider a laser beam whose input profile is
$I_{0} \neq 0 $ inside a simply connected domain $G$ and $I_{0} =
0$ on the smooth boundary $\Gamma$ of $G$ and outside it. Because
of functional relation $I_{0} = I_{0}(u)$, refractive index is
constant on $\Gamma$ too. More specifically, we have
\begin{equation}
\label{w_bound} w (\lambda) \bar{w} (\lambda) = 4 u_{0}, \quad{}
\lambda \in \Gamma.
\end{equation}
Boundary $\Gamma$ is mapped on the $2 \sqrt{u_{0}}$-radius circle.
In virtue of equation~(\ref{w_bound}), for $\lambda \in \Gamma$
one can write $w = 2 \sqrt{u_{0}} \exp (i \theta)$. Then, if $w$
is assigned in such a way that, for instance, $w(0) = 0$, where $0
\in G$ is assumed, variation of argument of complex number $w$ is
\begin{equation*}
\simbon{\Delta}{\Gamma} \arg w = 2 \pi.
\end{equation*}
Under mentioned conditions, it can be demonstrated that $w$ is a
homeomorphic mapping of domain $G$ onto $2\sqrt{u_{0}}-$radius
disk $\Gamma\;'$\cite{Iwaniec}. As a consequence, mapping $w$
preserve the topology of domain $G$. It could be interesting to
note that if $w = w(\lambda, z)$ ``evolves'' along $z$ according
to equation~(\ref{z_dependence}) and this evolution is compatible
with equation~(\ref{S_elliptic}), inverse mapping (it exists since
$w$ is homeomorphic) $\lambda = \lambda(w)$, with $w \in
\Gamma\;'$, provides us with the beam profile at any $z$.

Mapping $w = S_{x} - i S_{y} = \left(S_{x}, - S_{y} \right)$ can
be also regarded as a two-dimensional vector field on the
$\lambda-$plane associated with transverse components of the
wavefront normal unit vector. In the case $\varphi \equiv 0$, it
is has the simple form
\begin{equation}
\label{unit_vec} {\bf n} = \frac{\nabla S^{\ast}}{\left |\nabla
S^{\ast} \right |} = \left(\frac{S_{x}}{\sqrt{1+ 4 u}},
\frac{S_{y}}{\sqrt{1 + 4 u}}, \frac{1}{\sqrt{1+4u}} \right).
\end{equation}
From this point of view, mapping $w$ gives us information about
light-rays distribution around propagating direction.

\begin{figure}[h]
\centerline{\includegraphics[width=9cm]{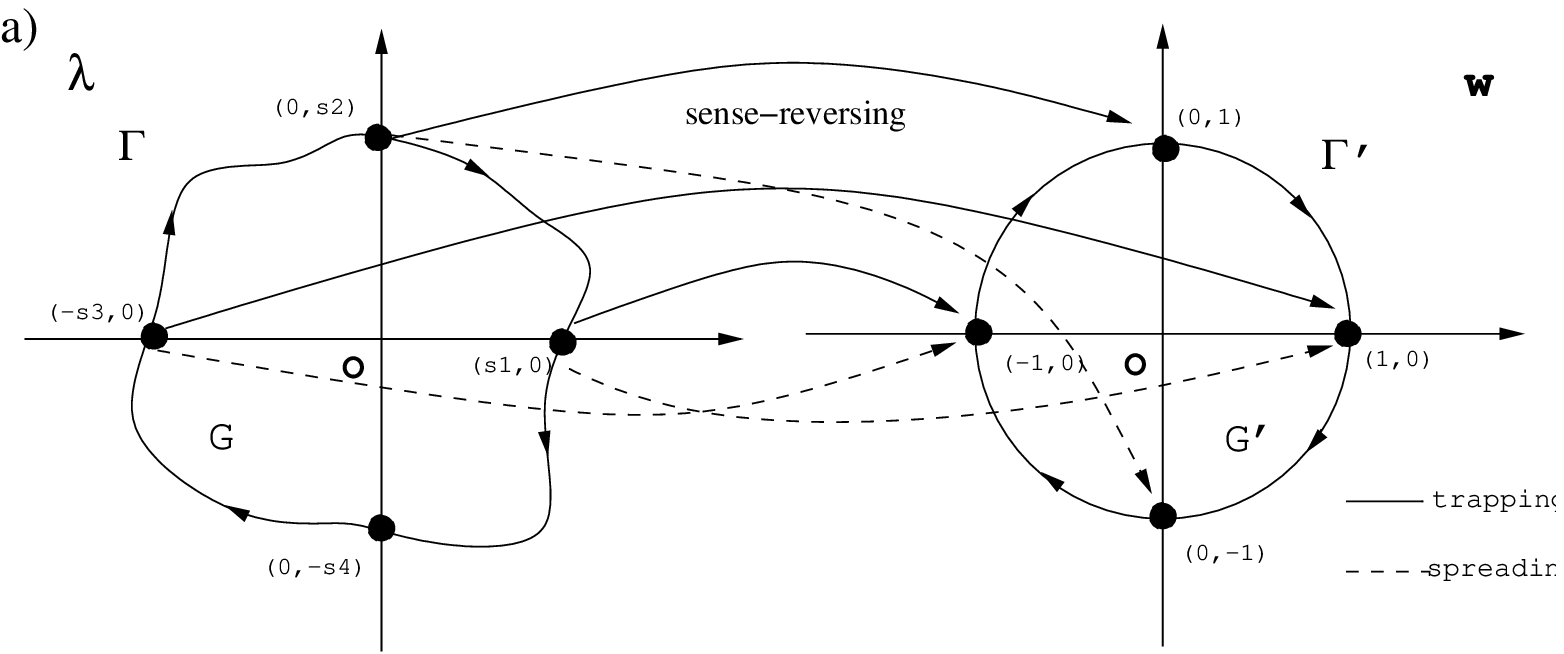}
\includegraphics[width=9cm]{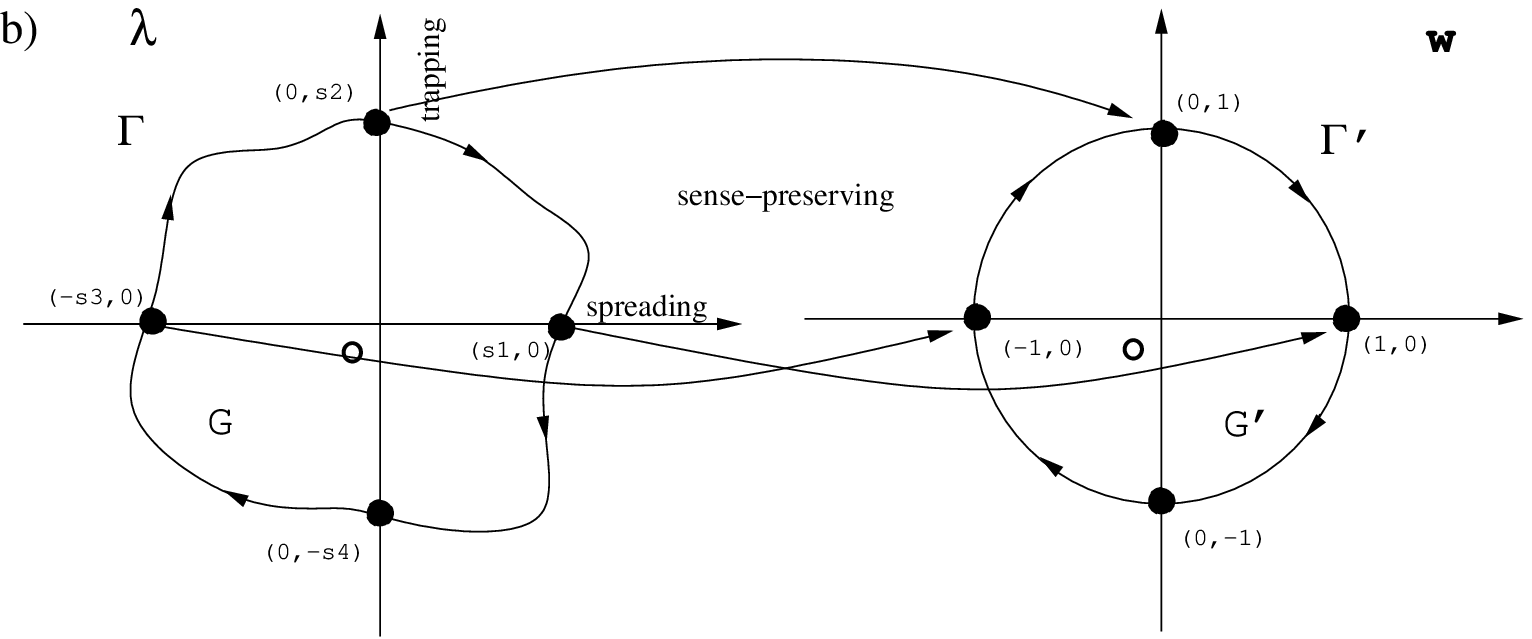}}
\caption{a) Sense-reversing mappings describe beams which tend to
be trapped (solid line) or spread (dashed line). b) If the mapping
is sense-preserving light beam spreads along $x-$direction and
tends to be trapped along $y-$direction. \label{figure1}}
\end{figure}

As specific example, let us consider a mapping as in
figure~\ref{figure1}a. Curves $\Gamma$ and $\Gamma'$ are oriented
leaving domain on the right hand side. Under conditions mentioned
above, there exists a homeomorphism $w$ of domain $G$ onto $G'$
acting in such a way that $w(s_{1},0) = (-1,0)$, $w(0,s_{2}) =
(0,1)$ and $w(-s_{3},0) = (1,0)$. Consequently, normal unit vector
to wave-front on the boundary $\Gamma$ points inside the domain
$G$ (recall that $y$-component of $w$ reverse $y$-component of
$\bf{n}$). This mapping allows the radiation to be `self-guided'
around propagating direction. Conversely, if $w(s_{1},0) = (1,0)$,
$w(0,s_{2}) = (0,-1)$ and $w(-s_{3},0) = (0,-1)$ radiation spreads
transversely far from $z-$axis. Note that in both cases
homeomorphism $w$ is sense-reversing. If we consider a
sense-preserving mapping such as, for instance, $w(s_{1},0) =
(1,0)$, $w(0,s_{2}) = (0,1)$ and $w(-s_{3},0) = (-1,0)$, the beam
spreads along $x-$direction and tends to be trapped along $y-$axis
(see figure~\ref{figure1}b). We emphasize, finally, that all of
these observations can be generalized to the case of arbitrary
$n-$connected domains.

\noindent Another class of solutions of
equation~(\ref{S_elliptic_complex}) can be obtained solving the
equation
\begin{equation}
\label{pre_reciprocal} b w_{\bar{\lambda}} + \bar{a}
\bar{w}_{\bar{\lambda}} = 0.
\end{equation}
Passing to reciprocal coordinates by inversion of the system
\begin{gather}
\label{recip_coordinates}
\begin{aligned}
\lambda &= \lambda(w,\bar{w}) \\
\bar{\lambda} &= \bar{\lambda} (w,\bar{w}),
\end{aligned}
\end{gather}
one converts equation~(\ref{pre_reciprocal}) into the form of the
{\em linear} Beltrami equation
\begin{equation}
\label{Beltrami_lin} \lambda_{\bar{w}} = \nu(w,\bar{w})
\lambda_{w},
\end{equation}
where $\nu (w,\bar{w}) = \bar{a}/b$ and, due to ellipticity,
$|\nu| < 1$. The advantage of the use of the reciprocal
transformation is that the Beltrami equation is just linear and
several properties of the analytic functions can be rigorously
generalized to its solutions (see e.g. Ref.~[\citenum{Vekua}]). An
interesting result is provided by the generalization of Liouville
theorem. Indeed, if $\lambda = \lambda(w,\bar{w})$ is
bounded on whole $w-$plane and satisfies linear Beltrami
equation~(\ref{Beltrami_lin}) it can be shown ({\em Vekua's
theorem}) that $\lambda (w,\bar{w}) \equiv
\textup{constant}$\cite{VekuaTh}. Of course, for the constant
solution, mapping from $w-$to$-\lambda-$plane is singular and
reciprocal transformation~(\ref{recip_coordinates}) is not
defined. Then, any non-trivial solution of
equation~(\ref{Beltrami_lin}) must be singular somewhere on the
complex plane and different type of singularities can occur, such
as poles, essential singularities, singularities of the
derivatives etc. A complete analytical and geometrical
classification of them will be done
elsewhere~\cite{InPreparation}.

For sake of simplicity, let us consider a solution
$\lambda(w,\bar{w})$ having a simple pole at $w = 0$. As
discussed above, one can always consider a homeomorphism from $D'
= \mathbb{C} \setminus D_{\epsilon}$ to $D_{R}$ where
$D_{\epsilon}$ is a disk of arbitrarily small radius $\epsilon$ on
the $w-$plane and $D_{R}$ is the $R$-radius disk on the
$\lambda-$plane, mapping the boundary of $D_{\epsilon}$ on the
boundary of $D_{R}$. Inverse mapping $w : \; D_{R} \to D'$
constructed in such a way can be used to describe a beam
``confined''around $z-$axis. Indeed, light rays crossing the
boundary of $D_{R}$ can settled down arbitrarily parallel to
$z-$axis as much as disk $D_{\epsilon}$ is small.

Coming back to nonlinear Beltrami equation~(\ref{Beltrami_nlin}),
we expect that for ``mild enough'' complex dilatations $\mu$,
Vekua's theorem still holds. In these cases, the only one bounded
solution on whole $\lambda-$plane is $w = \textup{constant}$. In
most of meaningful physical situations, intensity distribution on
the $\lambda-$plane, at certain $z$, goes to zero for $\lambda \to
\infty$, or equivalently, one can says that intensity vanishes
outside a big enough $R-$radius disk $D_{R}$. Thus, outside
$D_{R}$, refractive index assumes a constant value $u = u_{0}$ and
the solutions of eikonal equation~(\ref{eikonal}) is
\begin{equation}
S = c_{0} x + c_{1} y + c_{3},
\end{equation}
where $c_{0}$, $c_{1}$ and $c_{3}$ are constants and the condition
$c_{0}^{2} + c_{1}^{2} = 4 u_{0}$ holds. For a paraxial beam we
have $c_{0} = c_{1} = 0$. By a consequence $w = S_{x} - i S_{y} =
c_{0} - i c_{1} = 0$ and, of course it satisfies Beltrami equation
in $\mathbb{C} \setminus D_{R}$. In virtue of Vekua's theorem, the
only one bounded solution is $w \equiv 0$. Then, any non-trivial
solution must be singular somewhere on the plane. In our example,
wavefront is approximately plane for $\lambda \in \mathbb{C}
\setminus D_{R}$ and possesses singularity inside $D_{R}$.

\subsection{Minimal surfaces: the helicoid example.}
Standard approach to calculate solutions of Beltrami equation is
based on its reduction to a two-dimensional integral
equation~\cite{Vekua,Tricomi} which is amenable by using the
successive approximation method. In general, to find `explicit'
exact solutions is a quite challenging problem and successful
chances are critically depending on the form of the complex
dilatation. Incidentally, we note that solutions possessing
cylindrical symmetry such that
\begin{equation}
\label{cylindrical_sol} S = S(r), \quad{} u = u(r), \quad{} r =
\sqrt{x^{2} + y^{2}},
\end{equation}
are not compatible with intensity law. It is straightforward to
verify that equations~(\ref{eikonal}) and~(\ref{I0}) along with
assumptions~(\ref{cylindrical_sol}) imply that intensity depends
explicitly on $z-$axis distance $r$
\begin{equation}
\label{cylindrical_int} I_{0} = \frac{1}{2 c r \sqrt{u}},
\end{equation}
where $c$ is an arbitrary constant.

Here, we discuss solution of the
equation~(\ref{S_elliptic}) in connection with the minimal
surfaces. They are described by the following elliptic equation
(see e.g. Ref.~\citenum{Dubrovin})
\begin{equation}
\label{minimal} \left(1 + S_{y}^{2} \right) S_{xx} + \left (1 +
S_{x}^{2} \right) S_{yy} - 2 S_{x} S_{y} S_{xy} = 0.
\end{equation}
Now, we restrict ourselves to the class of solutions of
equation~(\ref{S_elliptic}) which are also harmonic, that is
\begin{equation}
\label{harmonic} S_{xx} + S_{yy} = 0.
\end{equation}
Using condition~(\ref{harmonic}) in equation~(\ref{S_elliptic}),
one gets the equation
\begin{equation}
\label{S_elliptic_min} S_{x}^{2} S_{xx} + S_{y}^{2} S_{yy} + 2
S_{x} S_{y} S_{xy} = 0,
\end{equation}
for any intensity law. It is straightforward to check
that equations~(\ref{S_elliptic_min}) and~(\ref{minimal}) coincide
for harmonic solutions. In other words, a class of solutions of
equation~(\ref{S_elliptic}), whatever is dependence $I_{0} =
I_{0}(u)$, is just given by the class of the harmonic minimal
surfaces. An interesting non-trivial example of solution is
\begin{equation}
\label{helicoid} S = K \arctan \left (\frac{y}{x} \right),
\end{equation}
where $K$ is an arbitrary constant. It is straightforward to check
that function $S$ in~(\ref{helicoid}) satisfies
equations~(\ref{minimal}), (\ref{harmonic})
and~(\ref{S_elliptic_min}) simultaneously. Equation of
corresponding wavefronts is
\begin{equation}
\label{helicoid_wavefronts} S^{\ast} \equiv z + K \arctan \left
(\frac{y}{x} \right) = \textup{const},
\end{equation}
where the constant $K$ is the ``pitch'' of the helicoid. We stress
that singular wavefronts of the form~(\ref{helicoid_wavefronts})
are well known in the theory of nonlinear singular optics and
their topological properties have important phenomenological
consequences  connected with description of screw wavefront
dislocations or so-called optical
vortices\cite{Vaughan,Baranova,Coullet} (see also
Ref.~[\citenum{Soskin1},\citenum{Soskin2}] and references
therein). Note that complex gradient associated with the helicoid
\begin{equation}
w = - i \frac{K}{\lambda}
\end{equation}
is analytic and it has a simple pole singularity at $\lambda = 0$.
Nevertheless, normal vector to the wavefront, whose components
coincides (up to a sign) with real and imaginary parts of $w$, is
not defined. Indeed, vector $w = (S_{x}, -S_{y})$ has no limit as
$\left (x, y\right) \to 0$. Figure~\ref{figure2} shows examples of
helicoidal wavefronts~(\ref{helicoid_wavefronts}) usually
parametrized as follows
\begin{equation}
x = v \cos t, \quad{}y = v \sin t, \quad{}z = K t.
\end{equation}
\begin{figure}[h]
\label{figure2}
\centerline{\includegraphics[width=15cm]{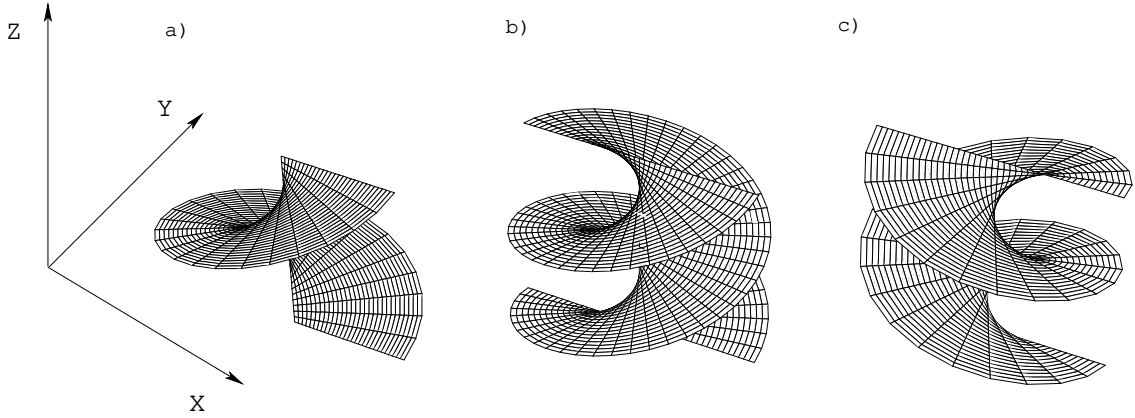}}
\caption{Helicoidal structures of wavefront around $z-$axis: a)
one-start right-screw ($K=1$); b) two-start right screw ($K=1$);
c) two-start left screw ($K=-1$). \label{figure2}}
\end{figure}
One-started helicoid shown in figure~\ref{figure2}a is obtained
for $0< v < + \infty$; two-started helicoids shown in
figure~\ref{figure2}b,c are obtained for $- \infty < v < +
\infty$. Refractive index
\begin{equation}
\label{helicoidal_u} u = \frac{K^{2}}{4 (x^{2}+ y^{2})}
\end{equation}
 has cylindrical symmetry around
$z-$axis and displays divergence at $\lambda = 0$. Thus, in most
of physical cases one has a divergence of intensity $I_{0}$. This
means that around $\lambda = 0$ geometrical optics approximation
fails and wave effects become relevant. In particular, necessary
condition for the existence of singular wavefronts is that
intensity vanishes where phase function is singular. Indeed, in
this region interference phenomenon is no more negligible and can
realize this condition.

Finally, we note that helicoidal wavefront is `stable', up to
$\omega^{-2 \nu}-$orders deformations, for a specific class of
first degree nonlocal perturbations. In this case, the
phase~(\ref{helicoid}) is defined up to an additive arbitrary
function of $z-$variable $\psi(z)$, such that $\psi' =
\varphi(z)$. Thus, using the phase expression $S = K \arctan
\left(y/x \right) + \psi(z)$ along with
relation~(\ref{helicoidal_u}) in
equations~(\ref{firstnonlocal_phi_hp})
and~(\ref{firstnonlocal_u_hp}), we conclude that the helicoid is
deformed for nonlocal harmonic data $\alpha_{1}$ and $\alpha_{2}$
satisfying simultaneously the following equations
\begin{gather}
\label{helicoidal_nonlocal_data}
\begin{aligned}
\left (x^{2} + y^{2} \right) \alpha_{1x} &= x \alpha_{1} + y
\alpha_{2} \\
- y \alpha_{1} + x \alpha_{2} &= \varphi(z) \left(x^{2} + y^{2}
\right).
\end{aligned}
\end{gather}
Trivial solutions $\alpha_{1} = \alpha_{2} = \varphi = 0$
corresponds to the local case discussed above. A simple
non-trivial solution $\alpha_{1} = - \gamma y$, $\alpha_{2} =
\gamma x$ and $\varphi = \gamma$, where $\gamma$ is an arbitrary
constant, provides us with the following wavefront (we remind that
to return to speed variables one has to substitute $\psi (z) \to
\psi(\omega^{-2\nu} z)$)
\begin{equation}
\label{helicoid_stretch} S^{\ast} \equiv z + \frac{K}{1 +
\omega^{-2 \nu} \gamma} \arctan \left(\frac{y}{x} \right) =
\textup{const}.
\end{equation}
Note that for $\gamma = 0$ equation~(\ref{helicoid_stretch})
coincides with the equation~(\ref{helicoid_wavefronts}). As a
consequence of nonlocal response the helicoid's pitch is
compressed if $\gamma > 0$ and stretched if $\gamma < 0$.


\section{Intensity law and nonlocal perturbations.}
\label{sec_symmmetry} Compatibility condition between dVN
hierarchy and the intensity law $I_{0} = I_{0}(u)$ is an as
challenging as intriguing issue. Indeed, due to the Beltrami
equation, it could establish a direct connection between
dispersionless integrable systems and an entire class of
quasiconformal functions parametrized by $z$. Unfortunately,
intensity law appears to be quite restrictive making the problem
of compatible nonlocal responses rather non-trivial one. This
section is devoted to demonstration of the existence of a suitable
intensity law such that systems~(\ref{phase})
and~(\ref{intensity}) are compatible for a third degree nonlocal
response.

To do that, we consider hydrodinamic type reductions of dVN
equation. They have been found using symmetry constraint of the
form $\nabla_{\bot}^{2}S = u_{x}$\cite{Bogdanov}. It can be shown
that dVN equation~(\ref{dVN}) is reduced to following
hydrodynamic-type system
\begin{align}
\label{hydro_system1} \left( \begin{array}{cc}
p_{1}  \\ p_{2} \\
\end{array} \right)_{y} &=
\left( \begin{array}{cccc}
0 & 1 \\
2 p_{1} - 1 & 2 p_{2}\\
\end{array} \right)
\left( \begin{array}{cc}
p_{1}  \\ p_{2} \\
\end{array} \right)_{x}\\
\left( \begin{array}{cc} p_{1}  \\ p_{2} \\
\end{array} \right)_{z} &=\left( \begin{array}{cc}
A_{11} & A_{12}  \\ A_{21} & A_{22} \\
\end{array} \right)\left( \begin{array}{cc} p_{1}  \\ p_{2} \\
\end{array} \right)_{x}
\end{align}
where
\begin{align*}
&A_{11} = 3 p_{1} \left( p_{1} -1 \right)   &A_{12} = 3 p_{2}, \\
&A_{21} = 3 p_{2} \left (2 p_{1}-1 \right)  &A_{22} = 3
  p_{1} \left(p_{1}-1 \right) + 6 p_{2}^{2}
\end{align*}
and $p_{1} := S_{x}$, $p_{2} := S_{y}$.

Now, looking for solutions such that $p_{2} = p_{2} (p_{1})$,
$p_{1}$ and $p_{2}$ are given in implicit form in terms of the
following algebraic system
\begin{align}
\label{dVN_solutions} &x + G' y + H' z - \Phi \left (p_{1} \right
)
= 0 \\
\label{p2_algebraic} &p_{2} = \frac{1}{2} \left [ \xi + \frac{2 c
- \log \left (\xi + \sqrt{1 + \xi^{2}} \right )}
{\sqrt{1 + \xi^{2}}} \right ]  \\
&\xi = p_{2} \pm \sqrt{p_{2}^{2} + 2 p_{1} -1} \nonumber
\end{align}
where $c$ is an arbitrary constant, $G = p_{2}(p_{1})$, $H =
p_{1}^{3} - \frac{3}{2} p_{1}^{2} + \frac{3}{2} p_{2}^{2}(p_{1})$
and `prime' means the derivative with respect to $p_{1}$.

Differentiating eikonal equation~(\ref{eikonal}) with respect to
$x$ and taking into account that $u_{x} = \nabla_{\bot}^{2}S$, one
gets
\begin{equation}
\label{eikonal_constraint} \nabla_{\bot}\left (- \frac{S_{x}}{2}
\right) \cdot \nabla_{\bot}S + \nabla_{\bot}^{2}S = 0.
\end{equation}
Comparing equation~(\ref{eikonal_constraint}) with intensity
equation~(\ref{I0}), one obtains the following nice relation among
intensity and $p_{1}-$component of the gradient
\begin{equation}
\label{intensity_symmetry} I_{0} = C\; e^{- \frac{p_{1}}{2}},
\end{equation}
where $C$ is an arbitrary real constant. Finally, eikonal equation
provides us with intensity law
\begin{equation}
\label{refractive_symmetry} u (I_{0}) = \left (\log
\frac{I_{0}}{C} \right)^{2} + \frac{1}{4} \left [ p_{2} \left (- 2
\log \frac{I_{0}}{C} \right) \right ]^{2},
\end{equation}
where last term in r.h.s. is given by algebraic
relation~(\ref{p2_algebraic}).

\section{Conclusions.}
\label{sec_conclusion} NNLS equation~(\ref{NNLS}) has been
proposed as the generalization of NLS equation in order to include
different degrees of nonlocal responses. Of course, just like
(2+1)D NLS equation, NNLS equation is not integrable. Particular
high frequency regimes discussed in this paper allows us to
investigate interplay among a wide class of nonlinear responses
and nonlocal perturbations free from interference effects. At
leading order non-trivial wavefronts exhibit singular vortex-type
behaviors. Moreover, singular phases can support propagation of
self-guided beams. Harmonic elliptic surfaces provide also with
stable beams with singular wavefronts such as the helicoid.
Moreover, there exists special first degree nonlocal perturbations
preserving helicoidal wavefronts. In the example considered, first
degree nonlocal perturbation produces a stretching or compression
of helicoid's pitch. A classifications of singularities and their
properties should provides us with a deeper understanding of this
phenomena and seems to be a promising subject of study.

On the other hand it has been shown that intensity law is
compatible with third degree nonlocal perturbations. In
particular, we note that new hydrodynamic-type reductions,
obtained by symmetry constraints approach can be very useful to
calculate new compatible intensity laws. This will stimulates our
efforts in that direction. Finally, we observe that although the
class of solutions in nonlocal perturbation regimes appears to be
not too large, example discussed above shows they are highly
non-trivial. Hence, in this regime, interesting and unexpected
phenomena could occur.


\acknowledgments B.K. and A.M. are supported in part by COFIN PRIN
``Sintesi'' 2004. A.M. would like to thank Professor M. Segev for
useful references concerning spatial solitons in nonlocal media.

\bibliographystyle{spiebib}

\end{document}